\documentclass[letterpaper, 10 pt, conference]{ieeeconf}  % Comment this line out if you need a4paper

\IEEEoverridecommandlockouts                              % This command is only needed if 
\usepackage{amssymb,amsmath,amstext,amsfonts}
\usepackage{mathtools}
\usepackage{bm}  % bold math anywhere in math mode and of greek letters too
\usepackage{siunitx}
\usepackage{physics}
\DeclareMathOperator*{\argmin}{arg\,min}

\usepackage{filecontents}
\usepackage{stmaryrd}

\usepackage{graphicx}
\graphicspath{%
	{Figures/}
	{Figures/TiKz/}
	{Figures/Ipe/}
	{Figures/XFig/}
	{Figures/Inkscape/}
	{Figures/Matlab/}
	{Figures/Scanned/}
	{Figures/Saved/}
}

\usepackage{subcaption}
\usepackage{cite}
\usepackage{url}
\usepackage{xcolor}

\let\labelindent\relax
\usepackage[inline]{enumitem}

\usepackage{algorithm,algpseudocode,float}
\usepackage{setspace} % Needed to fix line spacing inside the algorithms

\algnewcommand{\algorithmicand}{\textbf{ and }}
\algnewcommand{\algorithmicor}{\textbf{ or }}
\algnewcommand{\AlgAnd}{\algorithmicand}
\algnewcommand{\AlgOr}{\algorithmicor}
\newsavebox{\myparbox}
\newlength{\myparboxwidth}

\providecommand{\abs}[1]{\lvert#1\rvert}
\providecommand{\norm}[1]{\lVert#1\rVert}

\newcommand\Angle[1]{\setbox0=\hbox{$\mskip 7mu minus 4mu#1$}%
  \raise.21ex\hbox{$/$}\hskip-0.95ex\underline{\raise\dp0\hbox{\box0}}}

\ExplSyntaxOn

\NewDocumentCommand \vect { s o m }
 {
  \IfBooleanTF {#1}
   { \vectaux*{#3} }
   { \IfValueTF {#2} { \vectaux[#2]{#3} } { \vectaux{#3} } }
 }

\DeclarePairedDelimiterX \vectaux [1] {\lbrack} {\rbrack}
 { \, \dbacc_vect:n { #1 } \, }

\cs_new_protected:Npn \dbacc_vect:n #1
 {
  \seq_set_split:Nnn \l_tmpa_seq { , } { #1 }
  \seq_use:Nn \l_tmpa_seq { \enspace }
 }
\ExplSyntaxOff

\usepackage[firstpageonly=true]{draftwatermark}
\usepackage{hyperref}
\hypersetup{
    colorlinks=true,
    linkcolor=blue,
    filecolor=magenta,
    urlcolor=cyan,
    bookmarks=true,
}
\usepackage[noabbrev]{cleveref}  % automatically sort references
\Crefname{figure}{Fig.}{Figs.}

\title{\LARGE \bf An Adaptive Fuzzy Reinforcement Learning Cooperative Approach for the Autonomous Control of Flock Systems}

\author{Shuzheng Qu, Mohammed Abouheaf, Wail Gueaieb, and Davide Spinello%\thanks{*}% <-this % stops a space
\thanks{This work was partially supported by NSERC Grant~EGP~537568-2018.}%
\thanks{Shuzheng Qu, Mohammed Abouheaf, and Wail Gueaieb are with the School of Electrical Engineering \& Computer Science, while Davide Spinello is with the Department of Mechanical Engineering, University of Ottawa, Ottawa, Canada. E-mail:~\{fqu096,mabouhea,wgueaieb,dspinell\}@uottawa.ca.}%
}

\begin{document}

\maketitle
\thispagestyle{empty}
\pagestyle{empty}
%\bstctlcite{IEEEexample:BSTcontrol}  
%%%%%%%%%%%%%%%%%%%%%%%%%%%%%%%%%%%%%%%%%%%%%%%%%%%%%%%%%%%%%%%%%%%

\DraftwatermarkOptions{%
angle=0,
hpos=0.5\paperwidth,
vpos=0.97\paperheight,
fontsize=0.012\paperwidth,
color={[gray]{0.2}},
text={
  % Disclaimer as in
  % https://journals.ieeeauthorcenter.ieee.org/become-an-ieee-journal-author/publishing-ethics/guidelines-and-policies/post-publication-policies/
  % https://v2.sherpa.ac.uk/romeo/
  \newcommand{\thispaperdoi}{10.1109/ICRA48506.2021.9561204}
  \newcommand{\thispaperCopyrightYear}{2021}
  \parbox{0.99\textwidth}{This is the postscript version of the published paper. (doi: \href{http://dx.doi.org/\thispaperdoi}{\thispaperdoi})\\
    \copyright~\thispaperCopyrightYear~IEEE.  Personal use of this material is permitted.  Permission from IEEE must be obtained for all other uses, in any current or future media, including reprinting/republishing this material for advertising or promotional purposes, creating new collective works, for resale or redistribution to servers or lists, or reuse of any copyrighted component of this work in other works.}},
}

\begin{abstract}
The flock-guidance problem enjoys a challenging structure where multiple optimization objectives are solved simultaneously. This usually necessitates different control approaches to tackle various objectives, such as guidance, collision avoidance, and cohesion. The guidance schemes, in particular, have long suffered from complex tracking-error dynamics. Furthermore, techniques that are based on linear feedback strategies obtained at equilibrium conditions either may not hold or degrade when applied to uncertain dynamic environments. Pre-tuned fuzzy inference architectures lack robustness under such unmodeled conditions. This work introduces an adaptive distributed technique for the autonomous control of flock systems. Its relatively flexible structure is based on online fuzzy reinforcement learning schemes which simultaneously target a number of objectives; namely, following a leader, avoiding collision, and reaching a flock velocity consensus. In addition to its resilience in the face of dynamic disturbances, the algorithm does not require more than the agent position as a feedback signal. The effectiveness of the proposed method is validated with two simulation scenarios and benchmarked against a similar technique from the literature.
\end{abstract}

\section{Introduction}
The increasing complexity and diversity of modern robotic fields have raised new challenges and imposed limitations on various multi-agent robotic applications. For instance, many of such systems lack the autonomy required to control the collective behavior of a fleet of self-driving vehicles on a highway. Distributed control paradigms thrive in this domain~\cite{ICRA1,ICRA2}. They provide an alternative solution based on relaxing the inter-dependency requirements among the agents, such as the communication range of each agent, the amount of information required about its surrounding agents and the environment, etc. A key aspect about any distributed controller is its ability to optimize a set of simultaneously conflicting  local and team objectives.
%, like navigation, separation, and cohesion. 
The underlying control methods are often based on theories of feedback mechanisms and communication graphs~\cite{gu2008using,abouheaf2017flocking}.

%first attempts in this direction began by investigating the

The behavior of large numbers of interacting agents, such as flocks of flying birds and fish schools, is used in the study the cooperative dynamical systems~\cite{reynolds1987flocks,herd}. Sliding mode methods are employed to achieve  coordination of multi-agent systems in~\cite{ghasemi2014finite,li2018finite}. A neighbor-based tracking control scheme with distributed estimators is applied to advise distributed tracking strategies in~\cite{hu2010distributed}. Relevant control approaches relied on the existence of a virtual leader or a moving target in mobile sensor networks~\cite{5346069,4782025}. A common shortcoming to all these techniques is their dependence on the prior knowledge of a system model that is free of unstructured uncertainties. A distributed adaptive strategy that is based on a backstepping scheme is adopted to solve a consensus control problem in~\cite{Yang18}. The tracking problem of fractional-order multi-agent systems is addressed in~\cite{Li20} using adaptive strategies with neural network approximators.

Reinforcement Learning (RL) is a machine learning tool that allows agents to learn the best strategies to solve certain problems upon interacting with unknown environments~\cite{sut92,akinbulire2017reinforcement}. They do so by tracing the usefulness of being at a certain state while following some action in order maximize a cumulative reward criterion to eventually reach a target state. RL techniques offer flexible adaptations while interacting within complex environments~\cite{4445757}. They are usually implemented using either a Value Iteration or a Policy Iteration algorithm. A distributed fusion-based search strategy is developed to expedite the locomotion learning process for robots in~\cite{Cao19}. RL has helped with various robotic applications, such as unmanned flexible wing aircraft~\cite{AbouheafIETFWA,AbouheafTrans}, autonomous helicopters~\cite{abbeel2007application}, crowd aware robot navigation~\cite{ICRA3}, multi-robot predator avoidance~\cite{ICRA4}, guidance for biped humanoid robots~\cite{ICRA5}, multi-robot collision avoidance~\cite{ICRA6}, and efficient driving systems~\cite{article}. RL-based path-following control mechanisms for unmanned surface vehicles and unmanned aerial vehicles are presented in~\cite{woo2019deep,4554213}. To help approximate the unknown value function and the associated optimal strategy, actor-critic neural networks are usually utilized~\cite{sut92}. 
%The actor approximates the optimal strategy-to-follow while the critic reflects the value of being at a certain state while taking a particular action. 
The actor-critic structures have been adopted for cooperative control problems, such as graphical games and mobile sensor networks~\cite{8250107,5382498,AbouheafIETGraph}.
Despite their high potential in learning how to interact with ill-defined environments, RL schemes suffer from a major shortcoming stemming from the discrete (non-smooth) nature of their action space, which may lead to a ``jerky'' behavior when they are applied to control a real-world system. However, this is possible to solve by combining them with a fuzzy logic engine, taking advantage of its nonlinear approximation ability, to a get fuzzy RL algorithm with a continuous output space.

Fuzzy logic owes its popularity to its ability to incorporate human-like expertise in controlling complex systems in a model-free fashion.
%without the need of a prior mathematical model of the system's dynamics. 
%
It does so by defining the acquired meta knowledge about the system's functionality in terms of linguistic relations~\cite{ZADEH1965}. Fuzzy logic is employed to identify and control nonlinear dynamical systems in~\cite{lee2000identification}. Relevant fuzzy schemes have been applied in many processes, like temperature control~\cite{singhala2014temperature}, autonomous wheeled robot navigation~\cite{lin2005hierarchical}, and vehicle guidance~\cite{driankov2013fuzzy}.
They can be augmented with neural networks to automate their parameter-tuning process. For example, a neurofuzzy algorithm is designed to control a robotic manipulator in~\cite{er2003robust}. 
%An adaptive trajectory-tracking approach based on fuzzy systems is developed for nonholonomic mobile robots in~\cite{1624474}. 
Fuzzy schemes are used to guide mobile robots to follow a leader and to avoid obstacles in~\cite{innocenti2007multi,li2013design}. A collision avoidance approach based on an extended Takagi-Sugeno-Kang (TSK) inference system is proposed for a flock system in~\cite{abouheaf2017flocking}.

This work contributes an online fuzzy RL process that guides the motion of a flock. This is done while simultaneously compromising between competing local and team objectives. This architecture benefits novel model-free collision-avoidance and tracking mechanisms. It addresses concerns related to the dependence of the guidance control laws on prior knowledge about the dynamical models of the agents or the environment~\cite{AbouheafAut,AbouheafIJCNN,gu2008using}. Further, it enables online tuning of the fuzzy inference system without relying on a computationally exhaustive fuzzy Q-Learning search process. The rest of the paper is organized as follows: Section~\ref{problem_fromulation} formulates the problem and introduces a high level of the proposed control structure. The tracking, collision avoidance, and consensus control policies are detailed in Sections~\ref{track}, \ref{Sep}, and~\ref{Cons}, respectively. Section~\ref{Sim} presents and discusses the simulation results. Finally, Section~\ref{Conc} offers some concluding remarks.

\section{Problem Formulation and Control Structure}
\label{problem_fromulation}
We now formally define the cooperative control problem of a flock of $N$ agents communicating over a fully connected undirected graph. The goal is to make the flock follow a leader while satisfying a set of constraints. The leader does not have to be a physical agent. It may represent a virtual time-dependent signal or trajectory that would command the agents behavior. The problem is casted in a trajectory-tracking or pursuer-evader framework. The mobile agents are set to navigate in a 2D plane. The motion of each agent~$i$ is governed by
\begin{align*}
  \bm{p}_{k+1}^i & = \bm{p}_{k}^i + T \, {\bm{q}}_{k}^i
  &
  {\bm{q}}_{k+1}^i & = {\bm{q}}_{k}^i + T\, \bm{u}_k^i
\end{align*}
where $\bm{p}_{k}^i = [x^{i}_{k} \ y^{i}_{k}]^T \in \mathbb{R}^2$ and $\bm{q}_{k}^i = [v^{i,x}_{k} \ v^{i,y}_{k}]^T \in \mathbb{R}^2$ denote the position and linear velocity, respectively, of agent~$i \in \{1,2,\dots, N\}$ at discrete-time index $k\ \in \ \mathbb{N}$, $T$ is the sampling time, and $\bm{u}_k^i = [u^{i,x}_{k} , u^{i,y}_{k}]^T \in \mathbb{R}^2$ is the control signal commanding the acceleration in the x- and y-directions.
The control signals of all the agents are calculated by the proposed control algorithm, except the one for the leader $\ell \in \{1,\ldots,N\}$, which is generated by an independent command generator.
The location-measurements of the flock are set to be available for each agent.
% assuming a fully-connected undirected communication graph. 

% \subsection{Flock Objectives}
%
The aim of the flock is to satisfy the following objectives simultaneously
\begin{enumerate*}[before=\unskip{: }, itemjoin={{; }}, itemjoin*={{; and }}]
\item track the leader
\item keep a safe distance  among the agents
\item reach a velocity consensus.
\end{enumerate*}
The latter requirement involves an information layer among the agents. 
The agents have to manage the occasionally conflicting nature of such goals to reach the best possible compromise. They collectively search for a decision making policy that balances the local and team objectives. The local objective for each agent is to avoid colliding with its neighbors while staying close to the flock's focus area; while the team objectives involve reaching consensus on velocity for the flock and successfully following the leader.
Thus, the above three objectives can be formally described by the following respective relations for each pursuer~$i \in \{1,\ldots,N\} \setminus \{\ell\}$:
\begin{subequations}
\label{eq:objectives}
\begin{align}
  \label{eq:objectives:track-leader}
  & \lim_{k \to \infty} \norm{ \bm{p}_{k}^i - \bm{p}_{k}^\ell } = 0
  \\
  \label{eq:objectives:separation}
  & \lim_{k \to \infty} \abs{\zeta_k^i-\zeta_k^j} \geq d, ~\forall \zeta \in \{x,y\}, ~\forall j \in {\cal N}_i
  \\
  \label{eq:objectives:velocity-consensus}
  & \lim_{k \to \infty} v_k^{i,\zeta} = v_k^{i,*} %\text{, for some consensus speed $v_k^{i,*}$}
\end{align}
\end{subequations}
where ${\cal N}_i$ denotes the set of neighboring agents to agent~$i$, $d$ is a safety distance, $v_k^{i,*}$ is a consensus speed, and $\zeta \in \{x,y\}$ refers to either the $x$ or $y$ component.

To satisfy these objectives, the control command of each pursuer~$i \in \{1,\ldots,N\} \setminus \{\ell\}$ is formulated as an aggregate of three auxiliary signals
\begin{gather}
  {u}^{i,\zeta}_{k}
  = {u}_{t,k}^{i,\zeta}
  + {u}_{s,k}^{i,\zeta}
  + {u}_{c,k}^{i,\zeta}.
  \label{conttotal}
\end{gather}
The first term $u_{t,k}^{i,\zeta}$ is the tracking or navigation control signal. It is defined as a function of the agent's and leader's positions, $ u_{t,k}^{i,\zeta} \ =\ f^i_{t,\zeta}( \zeta_k^{i}, \zeta_k^{\ell})$, which handles the conflicts pertaining to these positions.
The separation or collision avoidance control of agent~$i$, $u_{s,k}^{i,\zeta}$, is applied to avoid colliding with agents $j$ in its neighborhood ${\cal N}_i$ in the $\zeta$ direction. It depends on the agent's position relative to its neighbors and is set as
\begin{align}
  {u}_{s,k}^{i,\zeta}
  &
    = f^i_{s,\zeta} (\zeta_k^{i} ,\zeta_k^{j})
    = \frac{\sum_{j \in {\cal N}_i} {u}_{s,k}^{ij,\zeta}}{\abs{{\cal N}_i}}
    \label{sep}
\end{align}
where $\abs{{\cal N}_i}$ denotes the cardinality of ${\cal N}_i$ and ${u}_{s,k}^{ij,\zeta}$ refers to the opinion or partial separation control signal taken by agent~$i$ due to each agent~$j\in{\cal N}_i$.
Finally, the synchronization goal is to guide all the pursuers to reach consensus on a common velocity. This is achieved by a consensus protocol $u_{c,k}^{i,\zeta} = f^i_{c,\zeta} (v_k^{i,\zeta} ,v_k^{j,\zeta})$, $\forall j \in {\cal N}_i$, which is defined in terms of the velocities of each agent and its neighbors.

This framework can be employed in many useful applications, including UAV monitoring and surveillance, area coverage, and mobile sensor networks, to name a few. This work does not consider the limitations imposed by the time-varying graph typologies.

% \noindent \textbf{\textit{Remark 1:}} Each agent \textit{i} makes a decision on how to react to its closeness to agent $j,j\in N_i$ (i.e., ${u}_{s,k}^{i,\zeta}$). Thus, there is an overall averaging decision made by agent \textit{i}  (i.e., $\frac{\sum ^{N_i-1}_{i = 1} {u}_{s,k}^{i,\zeta}}{N_i-1}$). Herein, each agent is following a leader and avoiding to collide with it as well.  Further, in this work it is assumed that each agent knows the positions of all other agents (i.e., $N_i \rightarrow N$). This averaging form is adaptable to other types of cooperative control problems.

% Mathematically speaking, the optimization problem aims to achieve the following goals
% \begin{enumerate}
% 	\item $\zeta_k^i-\zeta_k^\ell \rightarrow 0, \forall i$ as $k \rightarrow \infty$.
% 	\item $d\le\zeta_k^i-\zeta_k^j, \forall i,j$ as $k  \rightarrow \infty$ ($d$ is some safety distance).
% 	\item $v_k^{i,\zeta} \rightarrow v_k^{i,*}, \forall i$ as $k  \rightarrow \infty$ ($v_k^{i,*}$ is a consensus speed).
% \end{enumerate}

\section{Tracking Control Mechanism}
\label{track}
The section addresses the computation of the tracking control signal ${u}_{t,k}^{i,\zeta}$.
%
% articulates the tracking control solution for the cooperative control problem in hand using some ideas from the optimal control theory.

\subsection{Optimization Framework}
Let
$
\bm{E}^{i,\zeta}_{k} = [{\zeta}^i_{k} - {\zeta}^\ell_{k},\ {\zeta}^i_{k-1} - {\zeta}^\ell_{k-1},\ {\zeta}^i_{k-2} - {\zeta}^\ell_{k-2} ]^T \in \mathbb{R}^3
$
be an error signal vector corresponding to agent~$i$ at time index~$k$, representing a time window of the recent tracking error measurements between the agent's position and that of the leader. The choice of the number of tracking error instances
%
%include in $\bm{E}^{i,\zeta}_{k}$ 
%
depends on the complexity of the problem to solve. In our case, a time window of three components provided a satisfactory compromise between accuracy and computational complexity. In order for each agent~$i$ to measure the quality of its tracking control strategy $u^{i,\zeta}_{t,k}$, a performance index
$\chi_0^{i,\zeta}=\sum^{\infty}_{k=0}U^{i,\zeta}_{k} ( \bm{E}^{i,\zeta}_{k} ,u^{i,\zeta}_{t,k} )$
is proposed, where $U^{i,\zeta}_{k}$ is an objective function that is designed to be quadratic and convex
\begin{equation}
  \label{Con}
  U^{i,\zeta}_{k} ( \bm{E}^{i,\zeta}_{k} ,u^{i,\zeta}_{t,k} )
  = \dfrac{1}{2}
  \left[
    \bm{E}^{i,\zeta T}_{k} \bm{Q}^{i} \bm{E}^{i,\zeta}_{k} +{R}^{i} ( u^{i,\zeta}_{t,k} )^{2}
    \right]
\end{equation}
where $\bm{0} < \bm{Q}^i \in \mathbb{R}^{3 \times 3}$ and ${0} < R^i \in \mathbb{R}$ are some weighting factors corresponding to the tracking error vector and the control signal, respectively. When applied to matrices, the notations ``$>\bm{0}$'' and ``$\geq \bm{0}$'' denote positive definite and positive semi-definite matrices, respectively. 

Motivated by the structure of the performance index $\chi^{i,\zeta}$ and cost function~\eqref{Con}, a solving value function $V^{i,\zeta}$ is formulated as
\begin{gather*}
  V^{i,\zeta}( \bm{E}^{i,\zeta}_{k} ,u^{i,\zeta}_{t,k} )
  =\frac{1}{2}
  \left[
    \begin{array} {cc}
      {\bm{E}^{i,\zeta}_k}^T& {u^{i,\zeta}_{t,k}}
    \end{array}
  \right]
  \bm{H}^{i,\zeta}
  \left[
    \begin{array} {c}
      \bm{E}^{i,\zeta}_k\\ u^{i,\zeta}_{t,k}
    \end{array}
  \right]
  \\
  \text{such that, }
  \bm{H}^{i,\zeta}
  \equiv
  \begin{bmatrix*}[l]
    \boldsymbol{H}^{i,\zeta}_{\boldsymbol{E}^{i,\zeta}\boldsymbol{E}^{i,\zeta}}
    & \boldsymbol{H}^{i,\zeta}_{\boldsymbol{E}^{i,\zeta}{u_t^{i,\zeta}}}
    \\
    \boldsymbol{H}^{i,\zeta}_{{u_t^{i,\zeta}}\boldsymbol{E}^{i,\zeta}}
    &
    \boldsymbol{H}^{i,\zeta}_{{u_t^{i,\zeta}}  {u_t^{i,\zeta}}}
  \end{bmatrix*}
  \in \mathbb{R}^{4\times4}
\end{gather*}
where $\bm{H}^{i,\zeta} > 0$, $\boldsymbol{H}^{i,\zeta}_{{u_t^{i,\zeta}}  {u_t^{i,\zeta}}} \in \mathbb{R}$, and $\boldsymbol{H}^{i,\zeta}_{{u_t^{i,\zeta}}\boldsymbol{E}^{i,\zeta}} \in \mathbb{R}^{1 \times 3}$. 
This results in the following temporal difference (Bellman) equation
\begin{equation*}
V^{i,\zeta}( \bm{E}^{i,\zeta}_{k} ,u^{i,\zeta}_{t,k})=U_k^{i,\zeta}( \bm{E}^{i,\zeta}_{k} ,u^{i,\zeta}_{t,k} )+V^{i,\zeta}( \bm{E}^{i,\zeta}_{k+1} ,u^{i,\zeta}_{t,k+1} ).
\label{Bell}
\end{equation*}
Applying the optimality principle by taking  $\argmin_{u^{i,\zeta}_{t,k}}\left(V^{i,\zeta}( \bm{E}^{i,\zeta}_{k} ,u^{i,\zeta}_{t,k})\right)$ yields the optimal strategy-to-follow that is given by
\begin{equation}
  u^{i,\zeta(o)}_{t,k}
  = - \left({\boldsymbol{H}^{i,\zeta}_{{u_t^{i,\zeta}}  {u_t^{i,\zeta}}}}\right)^{-1} \boldsymbol{H}^{i,\zeta}_{{u_t^{i,\zeta}}\boldsymbol{E}^{i,\zeta}} \boldsymbol{E}^{i,\zeta}_{k}.
\label{opt}
\end{equation}
Employing the optimal policy in Bellman equation yields the Bellman optimality relation
\begin{multline}
  \label{Bello}
  V^{i,\zeta(o)}( \bm{E}^{i,\zeta}_{k} ,u^{i,\zeta(o)}_{t,k})
  =
  U_k^{i,\zeta}( \bm{E}^{i,\zeta}_{k} ,u^{i,\zeta(o)}_{t,k}) \\
  + V^{i,\zeta(o)}( \bm{E}^{i,\zeta}_{k+1} ,u^{i,\zeta(o)}_{t,k+1} )
\end{multline}
This equation is solved simultaneously by each agent~$i$ so that the agents can eventually converge to optimized tracking strategies.

%so that it can eventually converge to an optimized tracking policy.
% that would enable it to follow the leader. 

In this work, a two-step technique, known as Value Iteration, is applied to concurrently solve Bellman optimality~\eqref{Bello} using optimal policy~\eqref{opt} as follows:
%. This is achieved by iteratively evaluating the Bellman optimality and the optimal strategy concurrently,
%
\begin{equation*}
  \begin{split}
    V^{i,\zeta(r+1)}( \bm{E}^{i,\zeta}_{k} ,u^{i,\zeta}_{t,k})
    & = U_k^{i,\zeta}( \bm{E}^{i,\zeta}_{k} ,u^{i,\zeta(r)}_{t,k})
    \\
    & + V^{i,\zeta(r)}( \bm{E}^{i,\zeta}_{k+1} ,u^{i,\zeta(r)}_{t,k+1} )
    \\
    u^{i,\zeta(r+1)}_{t,k}
    & = - \left({\boldsymbol{H}^{i,\zeta(r)}_{{u_t^{i,\zeta}}  {u_t^{i,\zeta}}}}\right)^{-1} \boldsymbol{H}^{i,\zeta(r)}_{{u_t^{i,\zeta}}\boldsymbol{E}^{i,\zeta}} \boldsymbol{E}^{i,\zeta}_{k}
  \end{split}
\end{equation*}
till convergence~\cite{sut92}. This procedure leads to a converging nondecreasing sequence of solving value functions $0 \le V^{i,\zeta(0)}\le V^{i,\zeta(1)}\le \dots \le V^{i,\zeta(r)} \le\dots \le V^{i,\zeta(o)}$~\cite{abouctt}. 

\subsection{Neural Network Approximation}
An actor-critic approach is employed by each agent to execute the Value Iteration process. The actor is realized by a neural network approximator to estimate the optimal tracking strategy~\eqref{opt}. Another neural network is designed to implement the critic which approximates the optimal value function defined in~\eqref{Bello}. 

% The solution of the tracking optimization process is realized using neural network approximations to enable the solution of the Bellman optimality equation~\eqref{Bello} using the optimal policy~\eqref{opt}. Each agent employs two neural network approximations namely, the actor network is used to approximate the optimal tracking strategy and the critic network is set to asses the utility of this strategy. This is done by all agents simultaneously.    

The structures of the actor and critic approximators are motivated by the those of the optimal policy $u^{i,\zeta}_{t,k(o)}$ and optimal value function $V^{i,\zeta(o)}$. To this end, the optimal policy is approximated as $\hat u^{i,\zeta}_{t,k}= \bm{\omega}^{i,\zeta} \bm{E}^{i,\zeta}_{k}$, where $\bm{\omega}^{i,\zeta}\in \mathbb{R}^{1\times 3}$ is a row vector of the actor approximation weights. Similarly, the optimal value function $V^{i,\zeta(o)}$ is estimated by
\begin{gather*}
  \hat V^{i,\zeta}( \bm{E}^{i,\zeta}_{k} ,\hat u^{i,\zeta}_{t,k} )
  =
  \frac{1}{2}
  \left[
    \begin{array} {cc}
      \bm{E}^{{i,\zeta}^T}_k  & \hat u^{i,\zeta}_{t,k}
    \end{array}
  \right]
  \bm{\Omega}^{i,\zeta}
  \left[
    \begin{array} {c}
      \bm{E}^{i,\zeta}_k\\ \hat u^{i,\zeta}_{t,k}
    \end{array}
  \right]
  \\
  \text{such that, }
  \bm{\Omega}^{i,\zeta}
  \equiv
  \begin{bmatrix*}[l]
    \boldsymbol{\Omega}^{i,\zeta}_{\boldsymbol{E}^{i,\zeta}\boldsymbol{E}^{i,\zeta}}
    & \boldsymbol{\Omega}^{i,\zeta}_{\boldsymbol{E}^{i,\zeta}{\hat u_t^{i,\zeta}}}
    \\
    \boldsymbol{\Omega}^{i,\zeta}_{{\hat u_t^{i,\zeta}}\boldsymbol{E}^{i,\zeta}}
    &
    \boldsymbol{\Omega}^{i,\zeta}_{{\hat u_t^{i,\zeta}}  {\hat u_t^{i,\zeta}}}
  \end{bmatrix*}
  \in \mathbb{R}^{4\times4}
\end{gather*}
where $\bm{\Omega}^{i,\zeta} > 0$, $\boldsymbol{\Omega}^{i,\zeta}_{{\hat u_t^{i,\zeta}}  {\hat u_t^{i,\zeta}}} \in \mathbb{R}$, and $\boldsymbol{\Omega}^{i,\zeta}_{{\hat u_t^{i,\zeta}}\boldsymbol{E}^{i,\zeta}} \in \mathbb{R}^{1 \times 3}$.

A gradient descent approach is applied for the online training of the neural networks. The target approximation values of the optimal strategy and the optimal value function can be expressed as
$
\tilde u^{i,\zeta}_{t,k}=- \left({\boldsymbol{\Omega}^{i,\zeta}_{{\hat u_t^{i,\zeta}}  {\hat u_t^{i,\zeta}}}}\right)^{-1} \boldsymbol{\Omega}^{i,\zeta}_{{\hat u_t^{i,\zeta}}\boldsymbol{E}^{i,\zeta}} \boldsymbol{E}^{i,\zeta}_{k}
$
and
$
\tilde V_k^{i,\zeta}=U_k^{i,\zeta}( \bm{E}^{i,\zeta}_{k} ,\hat u^{i,\zeta}_{t,k})+\hat V^{i,\zeta}( \bm{E}^{i,\zeta}_{k+1} ,\hat u^{i,\zeta}_{t,k+1})
$,
respectively. As a result, the actor and critic approximation errors are
$\varepsilon^{i,\zeta(actor)}_{t,k}= \frac{1}{2}\left(\hat u^{i,\zeta}_{t,k}-\tilde u^{i,\zeta}_{t,k}\right)^2$
and
$\varepsilon^{i,\zeta(critic)}_{t,k}= \frac{1}{2}\left(\hat V^{i,\zeta}( \bm{E}^{i,\zeta}_{k} ,\hat u^{i,\zeta}_{t,k} )-\tilde V_k^{i,\zeta}\right)^2$, respectively.  Applying the gradient along the direction minimizing the errors yields the following weight update laws
\begin{align}
  \bm{\omega}^{i,\zeta(r+1)}
  & =
  \bm{\omega}^{i,\zeta(r)}
  - \rho_a\left(\varepsilon^{i,\zeta(actor)}_{t,k}  \bm{E}^{i,\zeta}_k\right)^{(r)}
   \label{RLact}
  \\
  \bm{\Omega}^{i,\zeta(r+1)}
  & =
  \bm{\Omega}^{i,\zeta(r)}
  - \rho_c\left(\varepsilon^{i,\zeta(critic)}_{t,k}  Z^{{i,\zeta}^T}_{t,k} Z^{i,\zeta}_{t,k}\right)^{(r)}
  \label{RLcrt}
\end{align}
where
$Z^{i,\zeta}_{t,k}= \vect{ \bm{E}^{{i,\zeta}^T}_k , \hat u^{i,\zeta}_{t,k} }$, $r$ is the iteration index of the weight update loop, and $0<\rho_a , \rho_c<1$ are the learning rates of the actor and critic, respectively.

\section{Separation Control Mechanism}
\label{Sep}
The separation control objective aims to prevent agents from colliding by imposing a safety distance between the agents. The RL protocol is designed such that it penalizes agents that are closer or further to each other than they should. Each agent~$i$ aggregates its decision based on its relative distance from its neighboring agents $j \in {\cal N}_i$, as described by~\eqref{sep}. The mechanism is realized through a zero-order Tagaki-Sugeno (TS) fuzzy logic inference engine with an online adaptation capability based on an actor-critic scheme. The synergistic integration of fuzzy logic and connectionist modeling theories has been exploited in the past to provide an approximate dynamic programming solution of the separation control problem~\cite{Jouf,akinbulire2017reinforcement}.
The merit of such adaptive fuzzy-RL scheme is its ability to approximate the highly nonlinear system's input-output relationship, using attractive-repulsive potential functions, on which it builds its optimized policy. Furthermore, it solves the granularity issue associated to the output space of classical RL approaches by converting it from a discrete action space to a continuous one.

\subsection{Zero-Order TS Fuzzy Inference System}
One of the most salient features of fuzzy logic is its ability to control complex systems characterized by dynamic uncertainties without the need for their precise mathematical models. It does so by incorporating human-like expertise, in the form of if-then rules, for the control of such ill-defined systems. 
An $A$-input single-output  zero-order TS fuzzy system employs $\cal P$ rules. Each rule~$p\in\{1,\ldots,{\cal P} \}$ is of the form
\begin{gather*}
  \text{If~}\theta_{1} \text{~is~} \Gamma^{(p)}_{m_{1}}, \dots , \text{~and~} \theta_{A} \text{~is~} \Gamma^{(p)}_{m_{A}}  \text{~Then~} u^{(p)} = \phi^{(p)},
\end{gather*}
where input $\theta_a$, $a\in \{ 1,\dots,A\}$, is an antecedent fuzzy variable, $\Gamma^{(p)}_{m_{a}}$ is a linguistic label associated to fuzzy set $m_a$, $u^{(p)}$ is a consequent fuzzy variable, and $\phi^{(p)} \in \mathbb{R}$ is a singleton. 
The firing strength of rule~$p$ is 
$\Psi ^{(p)} = \left. \qty[ \prod ^{A}_{a = 1} \eta ^{\Gamma^{(p)}_{m_a}}( \theta_{a})] \middle/  \sum ^{\cal P}_{p = 1}\left(\prod ^{A}_{a = 1} \eta ^{\Gamma^{(p)}_{m_a}}(\theta_{a})\right) \right.$
where 
$ \eta^{\Gamma^{(p)}_{m_a}}$ is a membership function associated to $\Gamma^{(p)}_{m_a}$. The fuzzy engine's inferenced output is computed through a defuzzification process as $u^{\text{Fuzzy}} = \sum ^{P}_{p=1} \Psi ^{(p)} \phi^{(p)}$.

\subsection{Adaptive Critics Fuzzy System}
A zero-order TS fuzzy system is adopted by each agent~$i$ to compute a separation signal ${u}_{s,k}^{ij,\zeta}$ to control its proximity to agent~$j$, $\forall j\in {\cal N}_i$. The fuzzy engine takes a single input fuzzy variable $\tilde{\zeta}^{ij}_k =\abs{\zeta^i_k-\zeta^j_k}-d$, where $d$ is the desired safety distance.  
With such a single-input single-output fuzzy system, rule~$p$ becomes of the form
\begin{gather*}
  \text{If~} \tilde{\zeta}^{ij}_k \text{~is~} \Gamma^{(p)}_{m_1} \text{~Then~} u^{ij,\zeta(p)}_{s,k}=\phi^{ij,\zeta(p)} .
\end{gather*}
The firing strength $\Psi_k^{ij,\zeta (p)}$ of rule~$p$ and the defuzzified output ${u}_{s,k}^{ij,\zeta}$ of the fuzzy system are computed as described above (with $A=1$).

In order to design an adaptive law to tune the consequent membership function $\phi^{ij,\zeta(p)}$ of each rule~$p$, the quality of the taken actions are assessed using a value function, which is approximated by a critic neural network 
${S}^{ij,\zeta}_{k} = \sum^{\cal P}_{p =1} \Psi_k^{ij,\zeta (p)} \Phi^{ij,\zeta(p)}$
with $\Phi^{ij,\zeta(p)}$ being the critic weight corresponding to rule~$p$.
A temporal difference equation can be formed in terms of the value function $S^{ij,\zeta}_k(\tilde{\zeta}^{ij}_k)$ and a reward function ${\cal R}^{ij,\zeta}_k$ as
$
S^{ij,\zeta}_k(\tilde{\zeta}^{ij}_k)={\cal R}^{ij,\zeta}_k+S^{ij,\zeta}_{k+1}(\tilde{\zeta}^{ij}_{k+1}).
$
The learning process of the actor and critic weights relies on such temporal difference equations.
The control objective here is to maximize instant rewards ${\cal R}^{ij,\zeta}_k$ based on the distance $\tilde{\zeta}^{ij}_k$, $\forall k$. 
Hence, the temporal difference error is represented by 
$
{\cal T}^{ij,\zeta}_k=S^{ij,\zeta}_k(\tilde{\zeta}^{ij}_k)-\left({\cal R}^{ij,\zeta}_k+S^{ij,\zeta}_{k+1}(\tilde{\zeta}^{ij}_{k+1})\right).
$
%
% The actor and critic weights are updated based on information about the temporal difference error ${\cal T}^{ij,\zeta}_k$ using the TS fuzzy system framework. This results in an adaptive fuzzy system where the consequences of the rules are continuously changing, until convergence is achieved, to cope with changes in the dynamic environment.
%
Applying a gradient-based descent approach, the update law of the actor weights is then
\begin{multline}
  {\phi^{ij,\zeta(p)}}^{(r+1)}  = {\phi ^{ij,\zeta(p)}}^{(r)} - \alpha_a \, \left(\text{sign}\left({\cal T}^{ij,\zeta}_k\right) \frac{\partial u^{ij,\zeta}_s}{\partial \phi^{ij,\zeta(p)}}\right)^{(r)}
  \\
  = {\phi^{ij,\zeta(p)}}^{(r)} - \alpha_a \, \left(\text{sign} \left({\cal T}^{ij,\zeta}_k\right) \Psi^{ij,\zeta (p)}\right)^{(r)}
  \label{FZact}
\end{multline}
where $0<\alpha_a<1$ is a learning rate.
Similarly, the adaptative law of the critic weights is derived from the temporal difference evaluation ${\cal T}^{ij,\zeta}_k$ as 
\begin{multline}
  {\Phi^{ij,\zeta(p)}}^{(r+1)}  = {\Phi^{ij,\zeta(p)}}^{(r)} - \alpha_c \, \left(\left({\cal T}^{ij,\zeta}_k\right) \frac{\partial {S}^{ij,\zeta}_{k}}{\partial \Phi^{ij,\zeta(p)}}\right)^{(r)}
  \\
  = {\Phi ^{ij,\zeta(p)}}^{(r)} - \alpha_c \, \left(\left({\cal T}^{ij,\zeta}_k\right) \Psi^{ij,\zeta (p)}\right)^{(r)}
  \label{FZcrt}
\end{multline}
for a learning rate $0<\alpha_c<1$. 
This adaptive critics structure is implemented by agent~$i$ in each direction $\zeta$ to provide an aggregate separation policy ${u}_{s,k}^{i,\zeta}$ defined by
\begin{align}
{u}_{s,k}^{i,\zeta}
&
= \frac{\displaystyle \sum_{j \in {\cal N}_i} \displaystyle \sum_{p=1}^{\cal P} 
	\Psi_k^{ij,\zeta (p)}  \phi^{ij,\zeta(p)}}{\abs{{\cal N}_i}}.
\label{sepo}
\end{align}
The resulting control process provides each agent with an online collision avoidance mechanism without the need to undergo many offline training episodes before applying the right decision~\cite{akinbulire2017reinforcement}.

\section{Consensus Control Mechanism}
\label{Cons}
The flock of agents is guided to reach a consensus on a common flock velocity using a communication layer designated by a graph topology. An undirected fully connected graph $\mathcal{G} = \{ \mathcal{N} , \mathcal{E} \}$ is adopted for this purpose, where $\mathcal{N} = \{\delta_i\}_{i=1,\ldots,\abs{\mathcal{N}}}$ is the set of nodes of cardinality $\abs{\mathcal{N}}$ and $\mathcal{E} = \{ (\delta_i,\delta_j) \in \mathcal{N}^2 \}$ is the set of edges representing the communication links between the agents~\cite{gu2008using}. We will denote the connectivity weights associated to every edge $(\delta_i,\delta_j) \in \mathcal{E}$ by $c_{ij}=c_{ji}$, $j\neq i$, with $c_{ii}=0$.
The local consensus protocol followed by each agent~$i$ is set to
\begin{equation}
  {u}^{i,\zeta}_{c,k} = - \sum _{j \in \mathcal{N}_{i}} c_{ij}(v^{i,\zeta}_{k} - v^{j,\zeta}_{k}).
\label{Consensus}
\end{equation}
Hence, the consensus control decisions for all agents can be presented collectively by $\bm{u}^{\zeta}_{c,k}= -\bm{L} \bm{v}^{\zeta}_{k}$, where  $ \boldsymbol{u}^{\zeta}_{c,k}=\vect{ {u}^{1,\zeta}_{c,k} , {u}^{2,\zeta}_{c,k} , \dots , {u}^{N,\zeta}_{c,k} }^T$, $ \boldsymbol{v}^{\zeta}_{k}=\vect{ {v}^{1,\zeta}_{k} , {v}^{2,\zeta}_{k} , \dots , {v}^{N,\zeta}_{k} }^T$, and $\bm{L}$ is the graph Laplacian. Let $\bm{C}=[c_{ij}] \in \mathbb{R}^{\abs{\mathcal{N}} \times \abs{\mathcal{N}}}$ be the graph's adjacency matrix and ${\bm C}^d=[c^d_{ii}] \in \mathbb{R}^{\abs{\mathcal{N}} \times \abs{\mathcal{N}}}$ be a square diagonal matrix where $c^d_{ii}=\sum _{j \in  \mathcal{N}_{i}} c_{ij}$. Then, $\bm{L} = {\bm C}^d - \bm{C}$.
The convergence speed of this process is governed by the second eigenvalue associated with Fiedler Eigenvector of the graph Laplacian $\bm{L}$~\cite{Saber}. As a result, the consensus relies on the graph topology as well as the cohesion features implicitly imposed by the tracking control law.

%As a result, the convergence is directly affected by the graph weights and topology.

\section{Simulation Results and Discussion}
\label{Sim}
% This section validates the performance of the Fuzzy-RL adaptive learning scheme using different scenarios. 

% \subsection{Simulation Scenarios}
Two simulation scenarios are set up to validate the performance of the proposed adaptive fuzzy-RL algorithm.
The first assigns a mobile agent as a leader and commands it to navigate in a circular trajectory defined by $x_k^\ell = 5\cos(0.03k)$, $y_k^\ell = 5\sin(0.03k)$. 20 other agents (followers) are then controlled by the fuzzy-RL algorithm to achieve the goals casted by~\eqref{eq:objectives}.
The agents' initial positions and velocities are randomly selected within the ranges of $\vect{-5,5}~\si{\m}$ and $\vect{0,1}~\si{\m/\s}$, respectively. The linear velocities in the x- and y-directions are bound to $\vect{-10,10}~\si{\m/\s}$. A reward function ${\cal R}^{ij,\zeta}_k$ is designed to give the highest value when the distance between the agents is equal to the target value~$d$. 
% while penalizing the actions violating this constraint
%
\begin{equation*}
  {\cal R}^{ij,\zeta}_k= 
  \begin{cases}
    - \tilde{\zeta}^{ij}_k, & \text{if } \tilde{\zeta}^{ij}_k > 0\\
    3, & \text{if } \tilde{\zeta}^{ij}_k = 0   \\
    \frac{3 \tilde{\zeta}^{ij}_k}{d}, & \text{if } \tilde{\zeta}^{ij}_k < 0    
  \end{cases}
\end{equation*}
The fuzzy logic engine is designed with five symmetric triangular membership functions as follows:
\begin{equation*}
  \eta ^{\Gamma_{m_1}}( \tilde{\zeta}^{ij}_k )= 
  \begin{cases}
    0, & \text{if } \tilde{\zeta}^{ij}_k<t_l \\
    \frac{\tilde{\zeta}^{ij}_k-t_l}{t_c-t_l}, & \text{if } t_l \leq  \tilde{\zeta}^{ij}_k \leq t_c   \\
    \frac{t_h-\tilde{\zeta}^{ij}_k}{t_h-t_c}, & \text{if } t_c \leq  \tilde{\zeta}^{ij}_k \leq t_h  \\  
    0, & \text{if } \tilde{\zeta}^{ij}_k > t_h  
  \end{cases}
\end{equation*}
The membership functions are centered around $t_c \in \{-6,-3,0,3,6\}$ with $t_l=t_h=1.5$. The input universe of the fuzzy logic controller is set to $(-7.5,7.5)$. These parameters reflect the action and state spaces for each agent, while the consequences of the fuzzy rules are decided online using the adaptive actor-critic structures. The remaining simulation parameters are taken as:
$\bm{Q}^{i}=I_{3\times3}$, ${R}^{i}=1$,
$\rho_c=10^{-7}$, $\alpha_c=0.05$, 
$\rho_a=10^{-2}$, $\alpha_a=0.1$, 
$d=\SI{2}{\meter}$, $T=\SI{0.1}{\second}$.
The neighborhood $\mathcal{N}_i$, for each follower~$i$, is defined to be $\mathcal{N}_i = \{\delta_j\}_{j=1,\ldots,N} \setminus \{\delta_i , \delta_\ell\}$. This makes $\abs{\mathcal{N}_i}=19$, since $N=21$.
%
% given in the following Table.
% \begin{table}[h]
%   \centering
%   \label{learn}
%   \caption{Simulation Parameters}
%   \begin{tabular}{cc|cc}
%     \hline 
%     Parameter	& Value & Parameter & Value \\ 
%     \hline 
%     $\bm{Q}^{i}$	& $I_{3\times3}$ & ${R}^{i}$ &  $1$ \\ 
%     \hline 
%     $\rho_c$ & $10^{-7}$ & $\rho_a$ & $10^{-2}$   \\ 
%     \hline 
%     $\alpha_a$ & $0.1$ & $\alpha_c$ & $0.05$ \\ 
%     \hline 
%     $T$	& $0.1\SI{}{\second}$ & $d$ & $2\SI{}{\meter}$ \\ 
%     \hline 
%   \end{tabular} 	
% \end{table}
%
In order to quantitatively assess the performance of the proposed algorithm, the following objective measures are adopted:
The average separation error for each follower~$i$ at time index~$k$ is measured as
$O_{s,k}^i = \qty[ \sum_{j \in \mathcal{N}_i} \qty( \norm{ \bm{p}_{k}^j - \bm{p}_{k}^i } -d ) ] \Big/ {\abs{\mathcal{N}_i}}$.
The overall average tracking error, separation error, and velocity are measured as
$O_{t,k}=\qty(\sum_{i \neq \ell} \norm{ \bm{p}_{k}^i - \bm{p}_{k}^\ell } ) \big/ {(N-1)}$,
$O_{s,k}=\qty(\sum_{i \neq \ell} O_{s,k}^i ) \big/ {(N-1)}$, and
$O_{v,k}=\qty(\sum_{i \neq \ell} \norm{ \bm{q}_{k}^i }  ) \big/ {(N-1)}$, respectively.

%\subsubsection{Scenario \# 1 - Circular Trajectory Tracking}

\begin{figure}
  \centering
  \begin{subfigure}[b]{2.5in}
    \centering
    \includegraphics[width=\columnwidth]{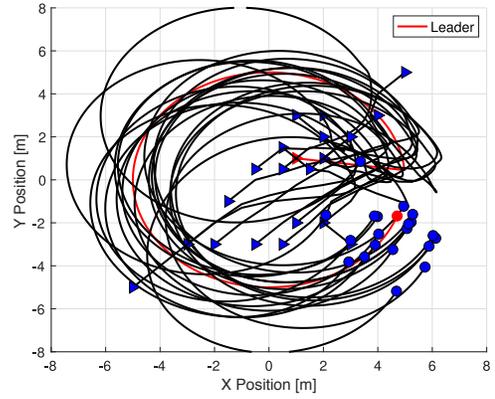}
    \caption[]%
    {Phase plane plot}    
    \label{fig:phase1}
  \end{subfigure}
  \\
  \begin{subfigure}[b]{2.5in}  
    \centering 
    \includegraphics[width=\columnwidth]{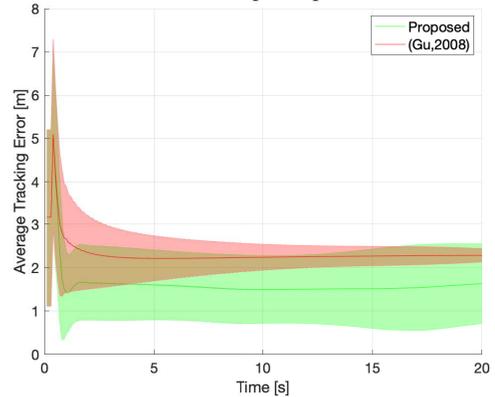}
    \caption[]%
    {Average tracking error~$O_{t,k}$} 
    \label{fig:trckerror1}
  \end{subfigure}
  \\
  \begin{subfigure}[b]{2.5in}   
    \centering 
    \includegraphics[width=\columnwidth]{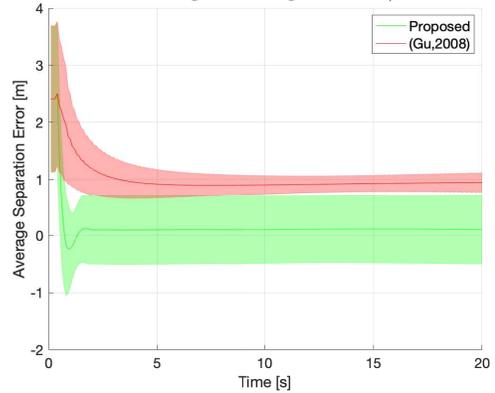}
    \caption[]%
    {Average separation error~$O_{s,k}$}   
    \label{fig:seperr1}
  \end{subfigure}
  \\
  \begin{subfigure}[b]{2.5in}   
    \centering 
    \includegraphics[width=\columnwidth]{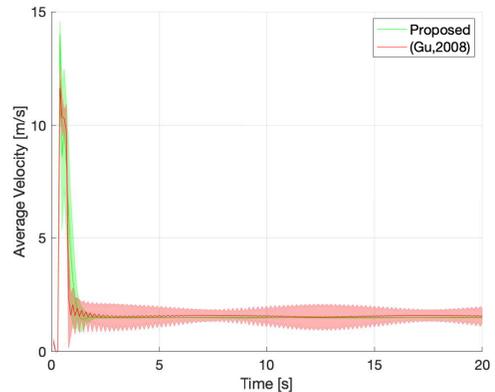}
    \caption[]%
    {Average follower velocity~$O_{v,k}$}    
    \label{fig:veloc1}
  \end{subfigure}
  \caption[]
  {Simulation results of Scenario~1} 
  \label{fig:main 1}
\end{figure}

The results are shown in Fig.~\ref{fig:main 1}. The performance is compared with a similar technique proposed by Gu et~al. in~\cite{gu2008using}. Fig.~\ref{fig:phase1} reveals the circular paths of the agents, where the start and end locations are indicated by a triangle and a circle, respectively. It is interesting to notice that with the adaptive fuzzy-RL method, not only do all the signals reach their respective steady states faster than with Gu's method, they do so within less than \SI{2}{\s}. Although this comes at the expense of a larger variation (standard deviation) for the average tracking and average separation errors, the agents reach a consensus on almost the exact same velocity with practically a nil standard deviation compared to Gu's method (Fig.~\ref{fig:veloc1}). The proposed method led to a significantly lower average tracking and average separation errors than Gu's controller. The average separation error with the adaptive fuzzy-RL algorithm rapidly converged to practically zero, outperforming Gu's algorithm whose error converged to \SI{1}{\m}. This means that while maintaining the competing objectives, the desired overall separation is achieved among the agents.

% As in Fig.~\ref{fig:phase1}, the agents follow a circular trajectory while making a compromise to keep each agent apart from the other. The overall average flock-performances related to the tracking error, separation error, and velocity of the mobile agents and their standard deviations at each instance $k$ are illustrated in Figs.~\ref{fig:trckerror1},~\ref{fig:seperr1}, and~\ref{fig:veloc1}, represented by red lines and light red shaded areas, respectively against those obtained by~\cite{gu2008using}, represented by green lines and light green shaded areas, respectively. As is evident form the tracking error plot in Fig.~\ref{fig:trckerror1}, the agents follow the leader keeping an almost constant overall average separation distance. This is well aligned with the ability of the adaptive learning scheme to preserve an overall average zero separation error as shown in Fig.~\ref{fig:seperr1}. Finally, the agents reach a consensus on a common flock velocity as shown in Fig.~\ref{fig:veloc1}. These figures reveal competitive performance compared to those obtained using~\cite{gu2008using}. For example, \cite{gu2008using}, aggressively separates the agents without paying attention to the overall average distraction. Consequently, its overall average tracking error is higher. Additionally a quicker flock consensus to a common speed is noted as per Fig.~\ref{fig:veloc1}. 

% \subsubsection{Scenario \# 2- Dynamic Missions}

The second simulation is designed to test the online ability of the fuzzy-RL controller to adapt to dynamic disturbances while the agents are in action. To that end, the flock starts off as in the first simulation, but with the following changes
\begin{enumerate*}[before=\unskip{: }, itemjoin={{; }}, itemjoin*={{; and }}]
\item at time $t=\SI{10}{\s}$, 4 of the 20 followers are decommissioned
\item at $t=\SI{20}{\s}$, the leader changes its trajectory from a circular to a linear trajectory defined by $v_k^{\ell,x}=1.7321$, $v_k^{\ell,y}=\SI{1}{\m/\s}$ (a linear motion with a heading of $\ang{30}$)
\item at $t=\SI{30}{\s}$, the safety distance $d$ is changed from 2 to \SI{2.5}{\m}.
\end{enumerate*}

\begin{figure}
  \centering
  \begin{subfigure}[b]{2.5in}
    \centering
    \includegraphics[width=\textwidth]{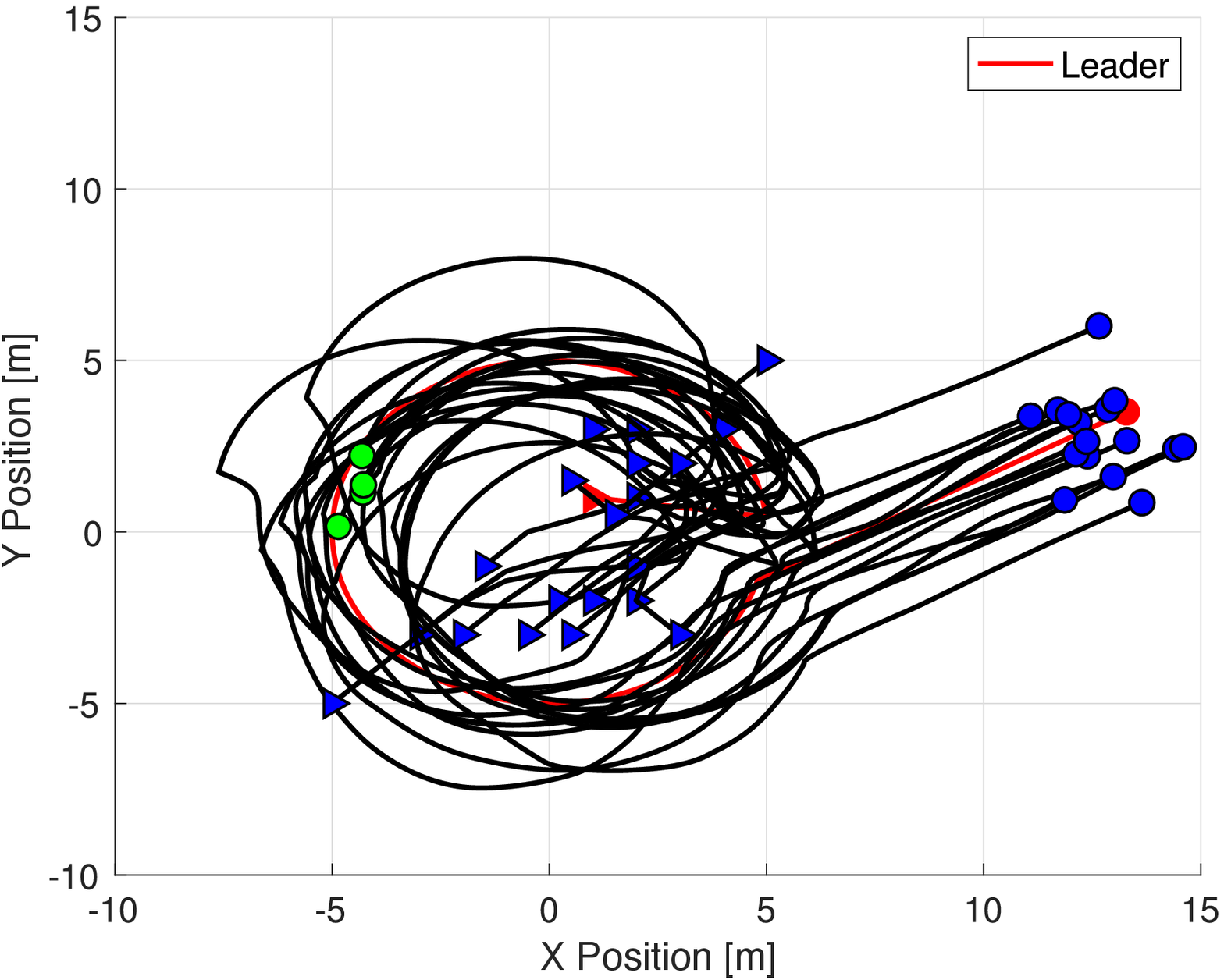}
    \caption[]%
    {Phase plane plot}  
    \label{fig:phase2}
  \end{subfigure}
  \\
  \begin{subfigure}[b]{2.5in}  
    \centering 
    \includegraphics[width=\textwidth]{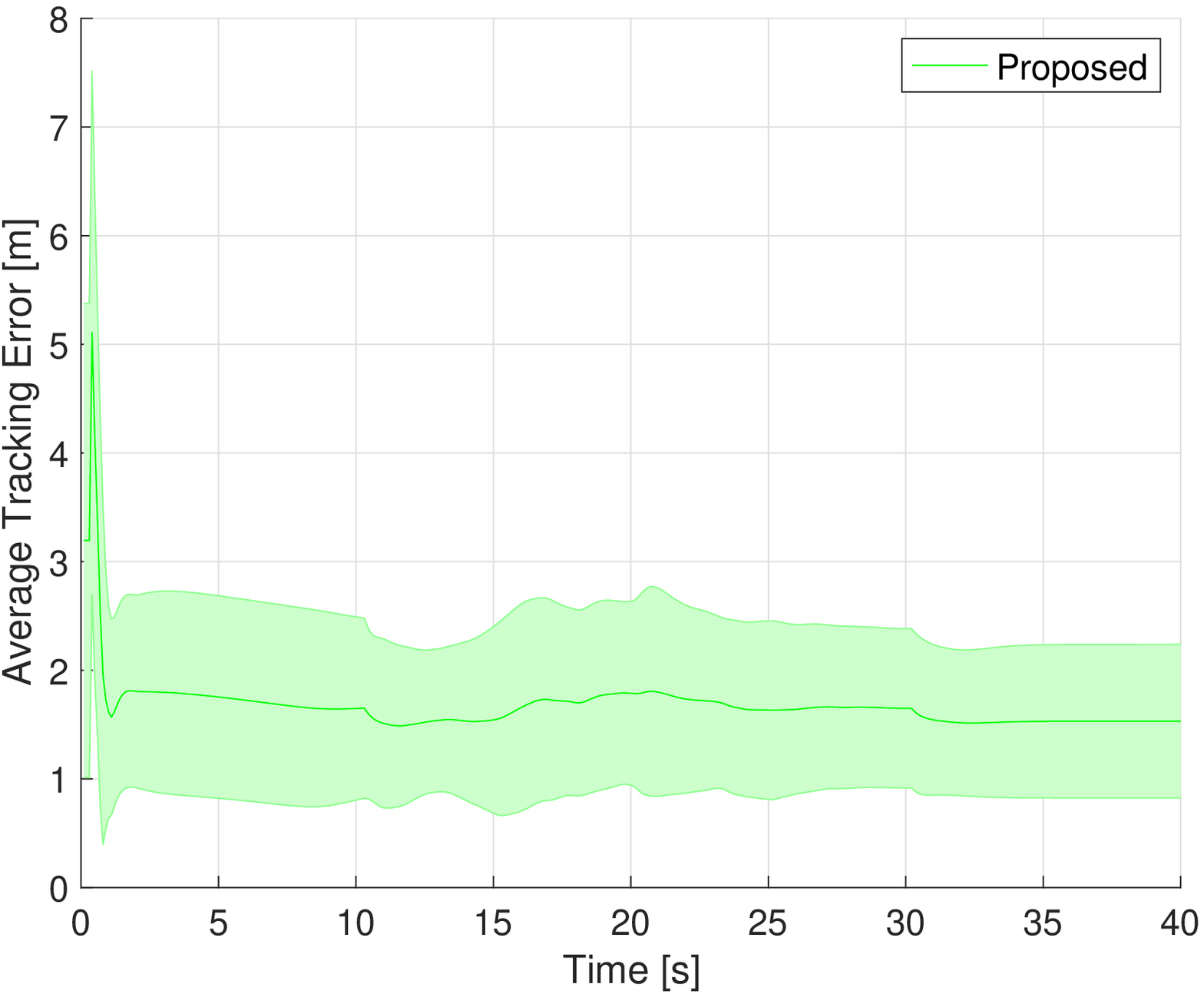}
    \caption[]%
    {Average tracking error~$O_{t,k}$}    
    \label{fig:trckerror2}
  \end{subfigure}
  \\
  \begin{subfigure}[b]{2.5in}   
    \centering 
    \includegraphics[width=\textwidth]{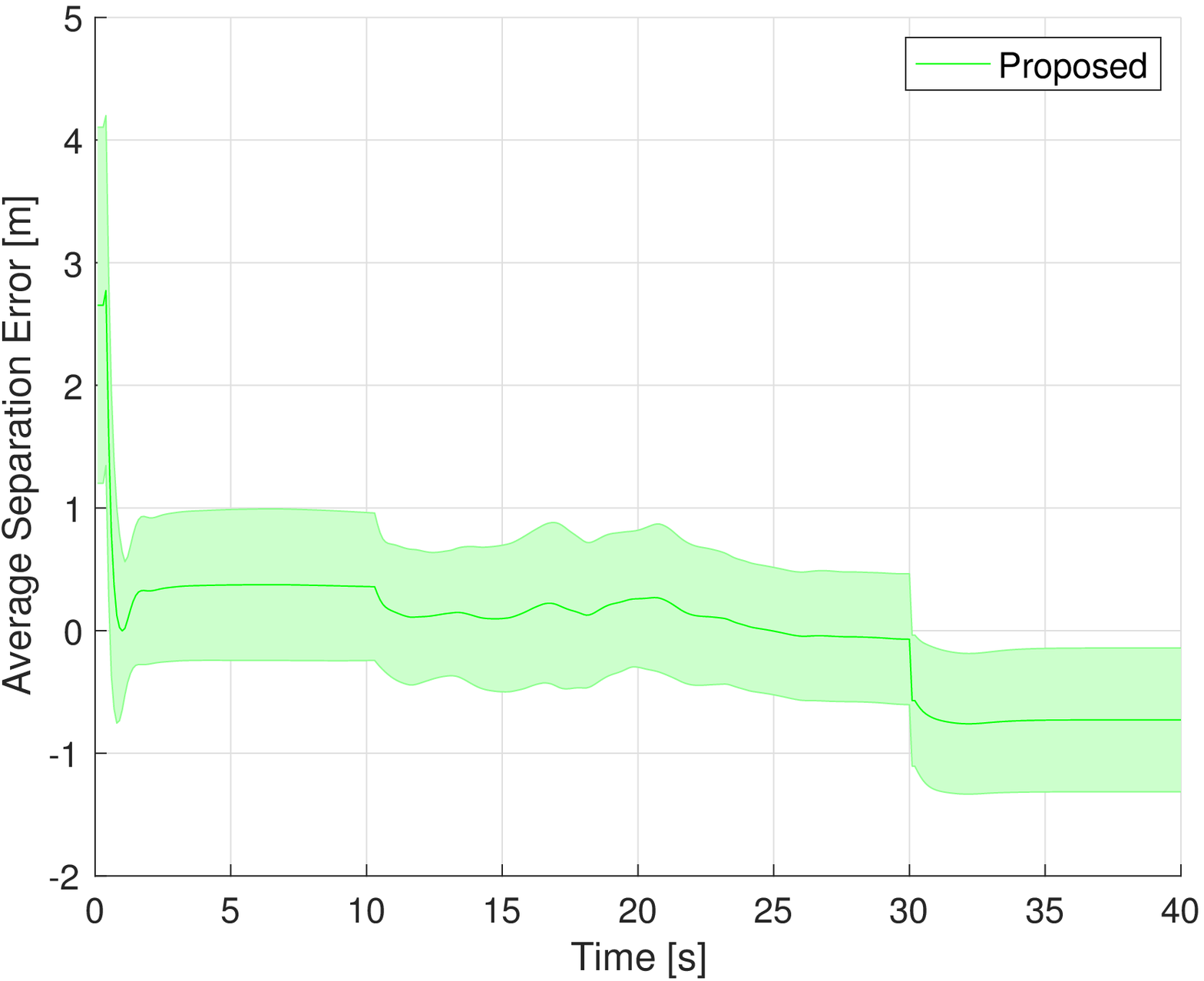}
    \caption[]%
    {Average separation error~$O_{s,k}$}    
    \label{fig:seperr2}
  \end{subfigure}
  \\
  \begin{subfigure}[b]{2.5in}   
    \centering 
    \includegraphics[width=\textwidth]{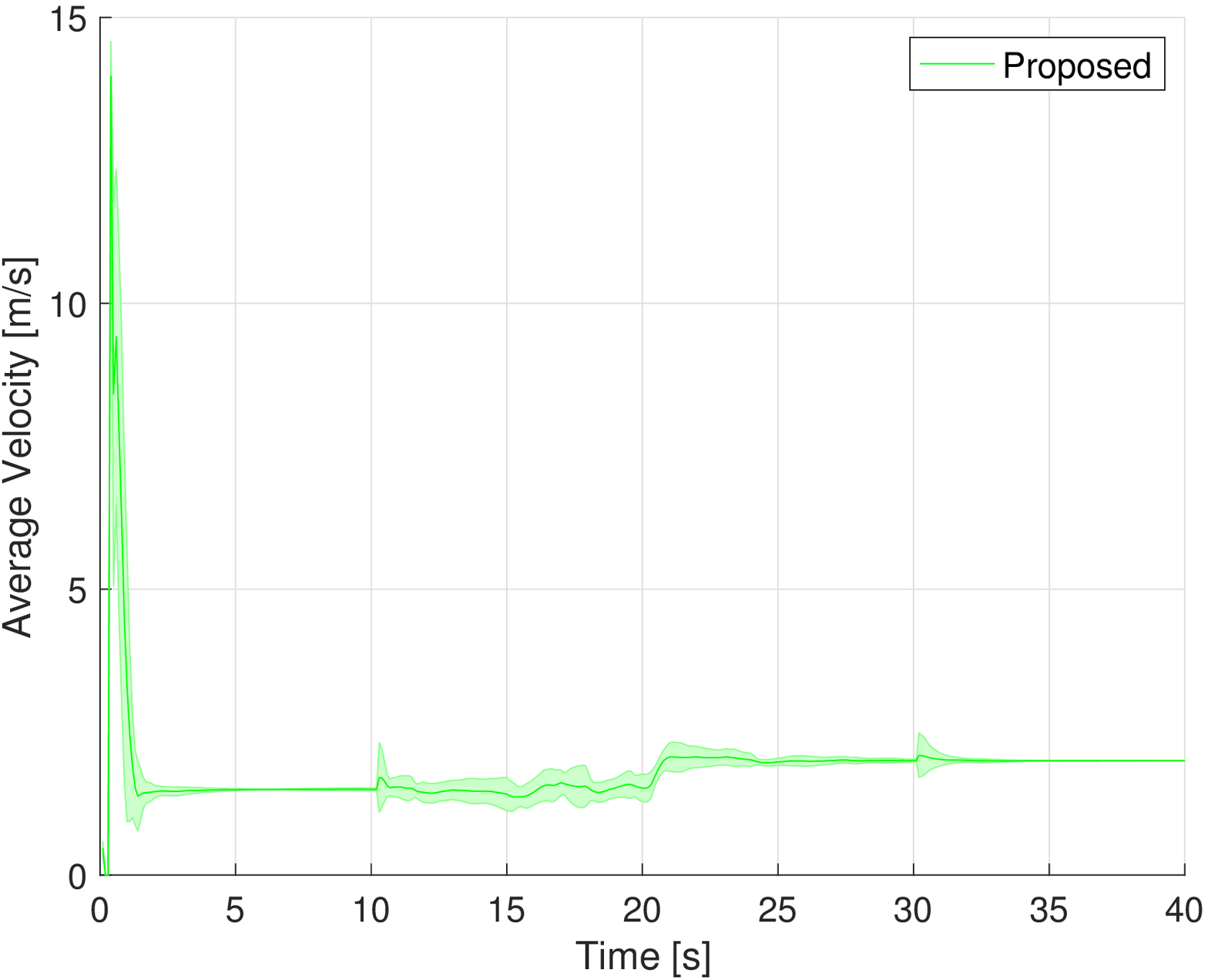}
    \caption[]%
    {Average follower velocity~$O_{v,k}$}    
    \label{fig:veloc2}
  \end{subfigure}
  \caption[]
  {Simulation results of Scenario~2} 
  \label{fig:main 2}
\end{figure}

The results of this scenario are depicted in Fig.~\ref{fig:main 2}. They are not compared with Gu's algorithm this time because it was not designed to be adaptive to dynamic variations. Its feedback policies are based on fixed control gains.
The 4 agents decommissioned at $t=\SI{10}{\s}$ are identified in green in Fig.~\ref{fig:phase2}. The figure demonstrates how the followers are able to globally track the leader regardless of the sudden change in trajectory or the number of active agents. The average tracking error, separation error, and velocity, rapidly converged right after each disturbance. Their standard deviations were not much affected by the changes in the simulation conditions. Nevertheless, the average separation error was the signal that is most affected by varying the safety distance from 2 to \SI{2.5}{\m} at $t=\SI{30}{\s}$. It goes from almost zero right before the change to about \SI{-0.75}{m} after it. On the other hand, Fig.~\ref{fig:veloc2} shows that the standard deviation of the consensus velocity converges to an insignificant value after each disturbance.

% The agents achieve successfully the different missions as shown in~Fig.~\ref{fig:phase2}. The tracking error, separation error, and velocity behaviors of the mobile agents are illustrated in Figs.~\ref{fig:trckerror2},~\ref{fig:seperr2}. These show the adaptability of the solution to dynamic missions.

\section{Conclusion}
\label{Conc}
An adaptive distributed online fuzzy reinforcement learning approach is proposed for the autonomous control of flock systems. It tries to simultaneously compromise three objectives
\begin{enumerate*}[before=\unskip{: }, itemjoin={{; }}, itemjoin*={{; and }}]
\item tracking a leader
\item avoiding collisions by maintaining a predefined safety distance between the neighboring agents
\item reaching a velocity consensus among the flock. 
\end{enumerate*}
The algorithm guides the flock towards a leader by applying a reinforcement learning scheme that is only dependent on the agent position feedback signals. The collision avoidance is realized by means of an adaptive fuzzy inference system that is based on its own reinforcement learning process. Finally, a graph-based protocol is employed to enable consensus on a common flock velocity.
In addition to its flexible and simple structure, the proposed algorithm is shown to possess a number of other salient features, such as fast convergence and online adaptability to dynamic disturbances. More specifically, it is proved to be relatively insensitive to the sudden changes in the leader's trajectory, number of active agents, and the target separation distance between the agents.  
The algorithm's superiority was demonstrated against a similar technique proposed in the literature.

% \clearpage

\bibliographystyle{IEEEtran}
\bibliography{Bib/mybibliography.bib}

\end{document}